\documentclass[twocolumn,showpacs,amsmath,prl,amssymb]{revtex4}

\usepackage{amsmath}
\usepackage{epsfig}
\usepackage{epsf}
\usepackage{array}
\usepackage{graphicx}

\begin{document}

\title{Suppression of superconductivity due to non-perturbative saddle points in
the nonlinear $\sigma$-model}
\author{D. A. Pesin}
 \affiliation{Department of Physics, University of Washington, Seattle, WA 98195, USA}

\author{A. V. Andreev}
 \affiliation{Department of Physics, University of Washington, Seattle, WA 98195, USA}

\begin{abstract}
We study superconductivity suppression due to thermal fluctuations in disordered
wires using the replica nonlinear $\sigma$-model ($NL\sigma M$). We show that in
addition to the thermal phase slips there is another type of fluctuations that
result in a finite resistivity. These fluctuations are described by saddle points
in $NL\sigma M$ and cannot be treated within the Ginzburg-Landau approach. The
contribution of such fluctuations to the wire resistivity is evaluated with
exponential accuracy. The magnetoresistance associated with this contribution is
negative.
\end{abstract}


\maketitle

Suppression of superconductivity in low dimensional samples has been studied for
a long time. In particular, in quasi-one-dimensional samples supercurrent decay
at temperature, $T$, below and near the critical temperature, $T_c$, is believed
to occur via thermally activated phase slips (TAPS). This idea was put forward by
Little~\cite{Little}, and the quantitative theory  was developed by Langer,
Ambegaokar, McCumber and Halperin (LAMH)~\cite{LA,McCH}. Within the LAMH theory
the wire resistance below $T_c$, $R$, has activated behavior, $R\propto
\exp(-\Delta F/T)$, where $\Delta F$ is the activation energy for a TAPS.

Most experimental results appear to be in good agreement with the LAMH
theory~\cite{Tinkham,Newbower}. However some discrepancies have also been
reported~\cite{Dynes,Chu}. In particular, the \emph{negative} magnetoresistance
observed in thin Pb  wires in Ref.~\cite{Dynes} has no explanation within the
LAMH theory.

From a theoretical viewpoint, the LAMH theory relies on the Ginzburg-Landau
description of thermal fluctuations resulting in finite resistivity.
\begin{figure}
\includegraphics[width=250 pt,bb=72 554 536 770]{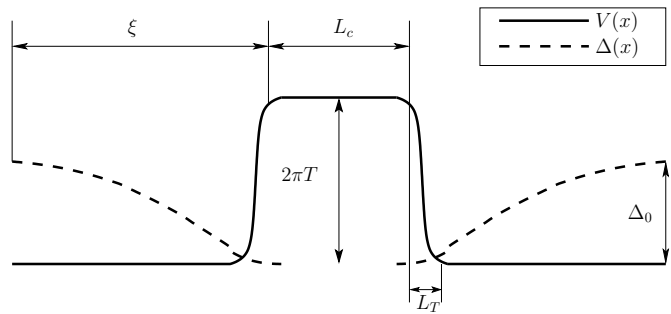}
\caption{Sketch of the electric potential, $V(x)$, (solid line) and
superconducting gap, $\Delta(x)$, (dashed line) dependence on the coordinate $x$
along the wire at the saddle point configuration.}\label{fig:gap}
\end{figure}
In this paper we study superconductivity suppression in disordered
wires using a more general description of a superconductor in terms
of the replica nonlinear $\sigma$-model ($NL\sigma
M$)~\cite{Finkelstein} that includes both the pairing and the
Coulomb interaction. This formalism describes the low energy physics
in terms of the order parameter, $\Delta$, the fluctuating electric
potential, $V$, and the $Q$-matrix that describes the fluctuating
electron Green function. Since the Ginzburg-Landau functional can be
obtained~\cite{Lerner} from the $NL\sigma M$ the LAMH optimal
fluctuation is automatically taken into account within this
formalism and appears as a saddle point of the $NL\sigma M$ action.

We show that due to the interplay of Coulomb interaction and disorder there exist
different saddle points of the $NL \sigma M$ action that provide alternative
channels for supercurrent decay. These fluctuations cannot be described within
the Ginzburg-Landau approach. They can be understood as follows. The interplay of
disorder and Coulomb interaction leads to soliton-like fluctuations of the
$Q$-matrix analogous to those first studied in Ref.~~\cite{Kamenev} in the
context of charge quantization in a single electron box with diffusive contacts.
In a disordered wire these solitons are accompanied~\cite{Andreev} by a change of
the \textit{imaginary} part of the electric potential by $2\pi T$ on the scale of
thermal diffusion length, $L_T=\sqrt{D/2\pi T}$, where $D$ is the diffusion
constant. Below $T_c$ the Coulomb and pairing interactions result in bound
soliton-antisoliton pairs with a typical spatial separation,  $L_c \sim \min
\{L_N, L_S \}$, with $L_N\sim \frac{e^2}{2\pi^2T}\ln\frac{e^2}{2\pi^2Td}$ and
$L_{S}\sim [\nu(T_c-T)\Delta^2_{0}A/T^2_c]^{-1}$, where $\nu$ is the density of
states at the Fermi level, $A$ is the wire cross section area, $\Delta_0$ is the
equilibrium value of the superconducting gap, $e$ is the electron charge, and $d$
is the transverse dimension of the wire. In the vicinity of the critical
temperature $L_c \gg L_T$, and the soliton and antisoliton within each pair are
well separated. We show below that superconductivity is suppressed in the middle
of the soliton-antisoliton pair, see Fig.~\ref{fig:gap}. This results in an
additional wire resistance
\begin{equation}\label{eq:resistance}
\delta R (T)= R_0 \exp\left[-2 G(L_T)-2\Delta F/T\right].
\end{equation}
Here $\Delta F/T \approx 0.35\, G(L_T) [(T_c-T)/T_c]^{3/2}$ is the activation
action of the LAMH phase slip, with $G(L_T)\equiv 4 \pi \nu D A /L_T \gg 1$ being
the dimensionless conductance of the wire segment of length $L_T$. Equation
(\ref{eq:resistance}) is valid for $G(L_T)^{-2/3} \ll (T_c-T)/T_c \ll
G(L_T)^{-1/2}$, where the first inequality corresponds to the Ginzburg criterion
and the second ensures the condition $L_c \gg L_T$.

Although rigorous evaluation of the pre-exponential factor, $R_0$, is rather
difficult and left for future work, we speculate that the interior of each
soliton-antisoliton pair acts as a segment of a normal metal, see discussion
below Eq.~(\ref{eq:solution_delta}). This reasoning leads to the estimate $R_0
\simeq R_N\frac{L^2_{c}}{L^2_T}$, where $R_N$ is the wire resistance in the
normal state.

The magnetoresistance associated with the contribution of these saddle points is
negative at low fields and is given by Eq.~(\ref{eq:magres}). This  may be
relevant to the negative magnetoresistance observed in Ref.~\cite{Dynes}.

We consider a long disordered wire with many transverse channels. The temperature
dependent coherence length $\xi(T)$ is assumed to exceed the transverse wire
dimension, $d$. The Thouless energy for the transverse motion, $E_T\equiv D/d^2$,
satisfies the condition $E_T\gg T$. In this regime the wire is described by a
one-dimensional $NL\sigma M$.

We write the replicated partition function for the system, $\langle
Z^n\rangle=\langle e^{-n\frac{\hat{H}}{T}}\rangle$, with $n$ being the number of
replicas, as an imaginary time functional integral over the fermions. Upon
disorder averaging, the $Q$-matrix (in the space of replicas and Matsubara
frequencies) is introduced to decouple the disorder-induced quartic interaction.
Its matrix elements are $4\times 4$ matrices in the space $S\otimes T$, given by
the product of spin, $S$, and time-reversal, $T$, spaces~\cite{Efetov83,Efetov}.
The auxiliary fields $V_a(x,\tau)$ and $ \Delta_a(x,\tau)$ decouple the Coulomb
and pairing interactions in the replica $a$ via Hubbard-Stratonovich
transformation. Proceeding in the usual way~\cite{Finkelstein}, one arrives at
the following form of the action for $Q$, $V$ and $\Delta$:
\begin{subequations}\label{eq:action}
\begin{eqnarray}
\label{eq:action_a}
\langle Z^n \rangle&=&\int {\cal{D}}[Q,V,\Delta]e^{-S_{Q}-S_{\Delta}-S_C},\\
\label{eq:action_b} S_Q&=&A\frac{\pi\nu}{2}\int dx\textrm{Tr}
\left[\frac{D}{4}(\nabla
Q)^2-(\hat{\varepsilon}+\hat{V}+i\hat{\Delta})Q\right]\nonumber\\
&&+ A \nu\int d\tau dx\sum_a V^2_a(x,\tau),\\
 \label{eq:action_c}
S_\Delta&=&A\int dxd\tau\sum_a\frac{|\Delta_a(x,\tau)|^2}{\lambda},\\
\label{eq:action_e} S_C&=&\!\frac{1}{2}\!\int \!d\tau dx dx'\sum_a
V_a(x,\tau)C(x,x')V_a(x',\tau).
\end{eqnarray}
\end{subequations}
Here $\textrm{Tr}$ denotes the trace over the replica, Matsubara and
$S\otimes T$ spaces. The matrices $\hat\varepsilon$, $\hat{V}$ and
$\hat{\Delta}$ have the following structure in the replica and
$S\otimes T$ spaces: $\hat\varepsilon=i\delta^{ab} \tau_3
\partial_\tau $, $\hat{V}=\delta^{ab}\tau_0V_a$,
$\hat{\Delta}=\delta^{ab}(\Delta_a\tau_+-\Delta_a^*\tau_-)$, where
$\tau_{\pm}=\frac{\tau_1\pm i\tau_2}{2}$ and $\tau_i$'s are defined as
$\tau_i=t_i\otimes\sigma_0$, with $\sigma_i$ and $t_i$ being the Pauli matrices in the
$S$ and $T$ spaces respectively. The positive constant $\lambda$ denotes the attraction
strength in the Cooper channel, and $C(x,x')$ describes the inverse effective Coulomb
interaction in the wire. In particular, for a homogeneous wire in the absence of a nearby
gate its Fourier component is $C(q)=1/e^2\ln{\frac{1}{q^2d^2}}$.

The saddle point equations are obtained by minimizing the action in
Eq.~(\ref{eq:action_a}) with respect to $\Delta$, $V$ and $Q$. For this purpose
one may neglect $S_C$ in~(\ref{eq:action}) if the number of channels in the wire
is sufficiently large, $e^2\nu A\gg 1$. This corresponds to the charge neutrality
limit~\cite{Andreev} and gives:
\begin{subequations}\label{eq:saddle}
  \begin{eqnarray}
    \label{eq:saddle_a}
    &&D\nabla(Q\nabla Q)-[\hat{\varepsilon}+\hat{V}+i\hat{\Delta},Q]=0,\\
    \label{eq:saddle_b}
    &&V_a-\frac{\pi}{4}\textrm{tr}\,Q^{aa}_{\tau\tau}(x)=0,\\
    \label{eq:saddle_c}
    &&\Delta_a+i\frac{\pi\nu\lambda}{4}\textrm{tr}[\tau_{-}Q^{aa}_{\tau\tau}(x)]=0,
  \end{eqnarray}
\end{subequations}
where $\textrm{tr}$ is the trace in the $S\otimes T$ space. Equation
(\ref{eq:saddle_a}) is the Usadel equation, Eq.~(\ref{eq:saddle_b})
represents the charge neutrality condition, and
Eq.~(\ref{eq:saddle_c}) plays the role of the BCS self-consistency
equation.

Before discussing the superconducting case, let us briefly describe solutions of
the saddle point equations (\ref{eq:saddle}) in a normal metal, $\Delta_a
=0$~\cite{Andreev}. Equations~(\ref{eq:saddle_a}) and~(\ref{eq:saddle_b}) possess
a set of degenerate stationary spatially uniform solutions, where
$Q^{ab}_{nm}=\delta^{ab}\delta_{n m}\tau_0\textrm{sgn}(\varepsilon_n+2\pi Tw_a)$,
$V_a=2\pi Tw_a$ with $w_a$ being an integer, $n, m$ being the Matsubara indices,
and $\varepsilon_n=\pi T(2n+1)$. The sum $4\sum_aw_a \equiv W$ (the factor $4$
here arises from the $4\times 4$ structure in the $S\otimes T$ space) defines the
trace of the $Q$-matrix, $\textrm{Tr} \, Q=2W$. Different $W$'s define
topological classes of $Q$, i.e. $Q$ matrices corresponding to the same $W$ can
be continuously rotated one into another in the replica and Matsubara spaces.
Thus in addition to the above spatially uniform saddle point configurations, the
action~(\ref{eq:action}) has spatially inhomogeneous finite-action solitons as
saddle configurations. These solitons connect different $Q$'s of the same
topological class, and correspond to both $Q$ and $V_a$ changing in space. These
nonuniform saddle points are crucial for the subsequent considerations.

Let us construct the soliton that connects $Q^{ab}_{nm}=\Lambda^{ab}_{nm}\equiv
\delta^{ab}\delta_{nm}\tau_0\textrm{sgn}\varepsilon_n$ at $x=-\infty$ and
$Q^{ab}_{n m}=\delta^{ab}\delta_{n m}\tau_0\textrm{sgn}(\varepsilon_n-2\pi Tw_a)$
at $x=\infty$, with $w^{1,2}=\mp 1$, and all the other $w$'s are zero. This
corresponds to gradual change in the electric potential in replicas $1$ and $2$
from zero at negative infinity to $\mp 2\pi iT$ at positive infinity. For
simplicity, we assume strong spin-orbit scattering and consider the $Q$-matrix
belonging to the symplectic ensemble~\cite{Efetov83}. In this case, the soliton
$Q$-matrix  can be parameterized by a single rotation angle
$\theta(x)$~\cite{Kamenev}:
\begin{equation}\label{eq:kink}
Q_{N}(x)=e^{-i\frac{\theta_i(x)}{2}\hat{u}\otimes \tau_{i}}\Lambda
e^{i\frac{\theta_i(x)}{2} \hat{u} \otimes \tau_{i}},
\end{equation}
where $\hat{u}$ is the generator of rotation between the Matsubara frequencies $\pi T$ in
replica $1$ and $-\pi T$ in replica $2$, $u^{ab}_{n m}=\delta^{a1}\delta^{b2}\delta_{n,
0}\delta_{m,-1}+\delta^{a2}\delta^{b1}\delta_{n,-1}\delta_{m,0}$. The index $i=0,1$
labels diffuson-like, $\tau_0$, and cooperon-like, $\tau_1$, rotations. Substitution of
Eq.~(\ref{eq:kink}) into (\ref{eq:saddle_b}) gives $V_{1,2}(x)=\mp \pi
T[1-\cos\theta(x)]$. Then Eq.~(\ref{eq:saddle_a}) gives
\begin{equation}\label{eq:sin-gordon}
\nabla^2\theta_i-\frac{1}{2L^2_T}\sin{2\theta_i}=0.
\end{equation}
This equation has a solution $\theta_i(x)=2\arctan(e^{x/L_T})$,
which gives for the electric potentials $V_{1,2}(x)=\mp \pi
T[1+\tanh(x/L_T)]$. The action (\ref{eq:action_b}) of this soliton
is $G(L_T)$. It can be shown that this saddle point is
stable~\cite{AADP}.

In a macroscopic system, solutions with nonzero $w_a$'s at
$x\rightarrow\pm\infty$ have infinite Coulomb action (\ref{eq:action_e}) and thus
are forbidden. Inhomogeneous saddle points appear as bound soliton-antisoliton
pairs with $V_a(x)$ appreciably different from zero only in the interior of the
pairs, see Fig.~\ref{fig:gap}. In the regime of interest, $G(L_T) \gg 1$, the
Coulomb action (\ref{eq:action_e}) does not affect the shape of the solitons, but
merely determines the typical soliton-antisoliton separation, $L_N$. The latter
can be estimated by equating the Coulomb action of a pair, $S_C \sim \frac{
2\pi^2TL_N}{e^2 \ln(L_N /d)}$,  to unity, resulting in $L_N\sim
\frac{e^2}{2\pi^2T}\ln\frac{e^2}{2\pi^2Td}$. We assume that $L_N\gg L_T$. This
assumption corresponds to $T\tau_{el}\ll
\frac{3}{2\pi^3}(\frac{e^2}{v_F})^2\ln^2\frac{e^2}{\pi^2Td}$ ($\tau_{el}$ is the
mean free time) and is always satisfied in the $NL\sigma M$ applicability region.
Since the soliton and antisoliton in a typical pair are well separated, the pair
action cost is $2G(L_T)$.

We now turn to the superconducting case. At $T>T_c$
Eq.~(\ref{eq:saddle_c}) implies $\Delta_a=0$, and the solutions of
Eqs.~(\ref{eq:saddle_a}), (\ref{eq:saddle_b}) remain the same as in
the normal case. Below $T_c$, in the region of interest, $T_c-T \ll
T_c$, a small static superconducting order parameter $\Delta_a$
appears. The saddle point solution, $Q_S$, can be expressed in terms
of the normal state solution (\ref{eq:kink}) as  $Q_S = U^\dagger
Q_N U$, where $U$ is a replica-diagonal rotation matrix in the
Gorkov-Nambu space, $U=\delta^{ab}U^a$, with
\begin{equation}\label{eq:Nambu}
U^{a}_{n m}=\delta_{n, m} \cos\frac{\phi^a_n}{2}-\tau_2 e^{-i\tau_3
\chi_a}\delta_{n,-(m+1)}
\textrm{sgn}(\varepsilon_n)\sin\frac{\phi^a_n}{2}.
\end{equation}
Here $\chi_a$ are the order parameter phases.

Solving Eq.~(\ref{eq:saddle_a}) using the ansatz (\ref{eq:Nambu}) with the
spatially uniform $Q_N=\Lambda$, and substituting the result into
(\ref{eq:saddle_c}) one recovers the Ginzburg-Landau equation for the order
parameter, which contains both the homogeneous BCS solution and the LAMH saddle
point.

Let us turn to the inhomogeneous $Q_N$, Eq.~(\ref{eq:kink}). In the
replicas $a \neq 1,2$ the situation is the same as for
$Q_N=\Lambda$. For $a=1,2$ substitution of the ansatz
(\ref{eq:Nambu}) into Eqs.~(\ref{eq:saddle}) leads to the following
equations for $\phi^a_n$, and $\Delta_a$,
\begin{subequations}\label{eq:delta}
\begin{eqnarray}
\label{eq:delta_a}
&&\!\!\!\!\!\!\!\!\!\!\!\frac{1+\cos\theta}{2}\nabla^2\phi^a_0\!+\!\nabla
\cos\theta \nabla\phi^a_0\!=\!\frac{\sin\phi_0}{L^2_T}
\!-\!\frac{2 \Delta_a}{D} \cos\phi^a_0,\\
\label{eq:delta_b} &&\!\!\!\!\!\!\!\!\!\!\!\nabla^2\phi^a_n=\frac{2
\varepsilon_n}{D}\sin\phi^a_n-
\frac{2 \Delta_a }{D}\cos\phi^a_n ,\\
\label{eq:delta_c}
 &&\!\!\!\!\!\!\!\!\!\!\!\frac{\Delta_a}{2\pi \nu \lambda T}=
\sum_{n>0}\sin\phi^a_n+\frac{1+\cos\theta}{2}\sin\phi^a_0.
\end{eqnarray}
\end{subequations}
Here we  assumed that there is no current in the wire and set the order parameter phase,
$\chi$, to zero. The additional equation for $\theta$ at $T_c-T \ll T_c$ may be
approximated by Eq.~(\ref{eq:sin-gordon}). Therefore below we consider $\theta$ in
Eqs.~(\ref{eq:delta}) to be the (approximate) solution of Eq.~(\ref{eq:sin-gordon})
corresponding to the soliton-antisoliton pair in the normal metal. Without loss of
generality we consider the soliton and antisoliton to be positioned at $x=\pm \zeta$ with
$\zeta \sim L_c \gg L_T$, where $L_c$ is the typical soliton-antisoliton separation in
the presence of superconductivity, estimated below.

The exact solution of the system  of equations (\ref{eq:delta}) is prohibitively
difficult. However, near the critical temperature the situation simplifies.
Outside the soliton-antisoliton pair, $|x| > \zeta$ the GL  approach may be used.
From the viewpoint of the GL description the presence of the soliton-antisoliton
pair imposes a boundary condition on the order parameter $\Delta_a$ at $x=\pm
\zeta$,
\begin{equation}\label{eq:bc}
    \nabla \ln \Delta_a (x)|_{x=\pm \zeta}=\pm \kappa,
\end{equation}
where $\kappa \sim L_T^{-1} \ll \xi(T)^{-1}$. The form of this
boundary condition can be obtained from the following
considerations. The effective ``critical temperature'', $T_c^*$, for
the appearance of a static order parameter, $\Delta_a$, is lowered
inside the soliton-antisoliton pair. Indeed,  for $\cos\theta=-1$
linearizing Eqs.~(\ref{eq:delta_b}) and~(\ref{eq:delta_c}) for a
uniform static $\Delta_a$ we obtain
\begin{equation}\label{eg:selfcon_old}
  \frac{1}{2\pi\nu\lambda}=T_c^* \sum_{n>0}\frac{1}
  { \varepsilon_n },
\end{equation}
where as usual the sum over Matsubara frequencies in the r.h.s. must be cut off at the
Debye frequency. The restriction $n>0$ in the summation arises because the last term in
Eq.~(\ref{eq:delta_c}) vanishes for $\cos \theta =-1$. Expressing the l.h.s. of
Eq.~(\ref{eg:selfcon_old}) in terms of $T_c$ we find $T_c^* =T_c /e^2$, where $e\approx
2.71$ is the base of the natural logarithm. Thus the regime of interest, $T_c -T \ll
T_c$, corresponds to the ``normal'' phase inside the soliton-antisoliton
pair~\cite{windcond}. Since $(T-T_c^*)/T_c^* \sim 1$ the effective correlation length
inside the pair is of order $L_T$, and the order parameter present outside the pair
penetrates into its interior to distances of order $L_T \ll \xi(T)$. This leads to the
boundary condition (\ref{eq:bc}) similar to the proximity effect for a normal metal in
contact with a superconductor~\cite{Abrikosov}. Then in the limit $L_T \ll \xi(T), h$ the
boundary condition (\ref{eq:bc}) may be replaced by $\Delta_a(\pm \zeta)=0$, and the
order parameter $\Delta_a(x)$ acquires the form,
\begin{equation}\label{eq:solution_delta}
|\Delta_a(x)| = \Delta_{0}\Theta(|x|-\zeta) \tanh{\frac{|x|-\zeta}{\sqrt{2}\xi(T)}},
\end{equation}
where $\Theta(x)$ is the step function.

Since the order parameter profile outside the soliton-antisoliton pair coincides with
that in the LAMH fluctuation in the zero current limit the action cost of such a
configuration is $S_0=2G(L_T)+2\Delta F /T$, where $\Delta F$ is the LAMH free energy
barrier  in the limit of zero bias current. The factor $2$ in front of $\Delta F$ arises
because the condensate suppression  occurs in two replicas. In this estimate we neglected
the Coulomb action and the condensation free energy in the interior of the pair, $S_\zeta
\sim \frac{2\pi^2T\zeta}{e^2 \ln(\zeta /d)} + \zeta A \nu(T_c-T)\Delta^2_{0}/T_c$. For a
typical fluctuation $S_\zeta \sim 1 \ll S_0$. This gives for a typical fluctuation size,
$L_c \sim \min \{L_N, L_S \}$, where $(L_S)^{-1}=\nu(T_c-T)\Delta^2_{0}A/T^2_c$.

With exponential accuracy the contribution of such fluctuations to the wire resistance is
determined by their action, which results in Eq.~(\ref{eq:resistance}). Evaluation of the
pre-exponential factor $R_0$ is a difficult task. It may in principle be accomplished
using linear response theory. This involves a cumbersome procedure of analytic
continuation of the Matsubara susceptibility to real frequencies and is outside the scope
of the present work. It is nevertheless plausible that the interior of the
soliton-antisoliton pair acts as a segment of a normal metal~\cite{snsjunction}, giving a
contribution $ \frac{h}{e^2G(L_T)}\frac{L_c}{L_T}$ to the wire resistance. The number of
soliton-antisoliton pairs in the wire can be estimated as $ \frac{L L_c}{L^2_T}e^{-S_0}$,
where the prefactor corresponds to $L/L_T$ ways to accommodate a soliton, and $L_c/L_T$
possibilities to put an antisoliton next to it. Following this argument one arrives at
the estimate $R_0\simeq R_N\frac{L^2_c}{L^2_T}$.

We now discuss the magnetic field dependence of resistance~(\ref{eq:resistance}).
It arises from the decrease of condensation energy $\Delta F$ due to magnetic
field, and an increase of the action of cooperon-like solitons.  In the presence
of a magnetic field, $H$, the $NL\sigma M$ action for a wire can be
obtained~\cite{Efetov} by changing $(\nabla Q)^2 \to (\nabla Q)^2-e^2 \langle
\mathbf{A}^2 \rangle [\tau_3,Q]^2$ in Eq.~(\ref{eq:action_b}), where $ \langle
\mathbf{A}^2 \rangle$ is the square of the vector potential, $\mathbf{A}$,
averaged over the wire cross-section (we assume that the magnetic length,
$l_H=\sqrt{\frac{1}{eH}}$, is much larger than the wire width, $d$). Each soliton
or antisoliton constituting a pair can be generated either by a diffuson-like or
a cooperon-like rotation, $i=0,1$ in Eq.~(\ref{eq:kink}). The diffuson is
insensitive to the magnetic field whereas the cooperon acquires a mass. As a
result, Eq.~(\ref{eq:sin-gordon}) for the cooperon rotation angle on the soliton
configuration becomes
\begin{equation}
\nabla^2\theta_c-\frac{1}{2}\left(\frac{1}{L^2_T}+
\frac{1}{L^2_H}\right)\sin{2\theta_c}=0,
\end{equation}
where $L^{-2}_H=e^2\langle \mathbf{A}^2\rangle \propto H^2$ depends on the specific
geometry, in particular for a rectangular wire of width $d$, and a magnetic field
perpendicular to it, $L^2_H=3 l_H^4/d^2$. The corresponding action change for a single
cooperon-like soliton is $ G(L_T)\left(\sqrt{1+L_T^2/L_H^2}-1\right)$. The change in the
condensation action, $\Delta F/T$, may be evaluated using the Ginzburg-Landau theory and
is much smaller, of the order of $G(L_T)\frac{L_T^2}{L_H^2}\sqrt{(T_c-T)/T_c}$. At
relatively weak fields, $L_H^2 \gtrsim G(L_T) L_T^2  \gg L_T^2$, it may be neglected
while the action change for the cooperon-like soliton can be significant. In this regime
we obtain~\cite{prefactor}
\begin{equation}\label{eq:magres}
  \delta R(H)\approx  \frac{\delta R (T)}{4} \left[ 1+ e^{ -G(L_T)
  L_T^2/2L_H^2}\right]^2.
\end{equation}
Here $\delta R(T)$ is given by Eq.~(\ref{eq:resistance}) and we took
into account that there are four ways to assemble a
soliton-antisoliton pair out of diffuson-like and cooperon-like
solitons. From Eq.~(\ref{eq:magres}) we see that the
resistance~(\ref{eq:resistance}) decreases with magnetic field.

In summary, we have considered superconductivity suppression in thin wires due to
non-perturbative saddle points in the $NL\sigma M$. We showed that their presence
leads to an additional (to the LAMH) contribution to the wire resistance below
$T_c$ and evaluated it with exponential accuracy.  This contribution is
suppressed by a magnetic field, which may correspond to the negative
magnetoresistance observed in Ref.~\cite{Dynes}.

We are grateful to I. Aleiner for numerous discussions that initiated this
project. This work was supported by the David and Lucille Packard Foundation.


\begin{thebibliography}{99}
\bibitem{windcond} Condensation at nonzero Matsubara frequencies, $\Delta(\tau) \propto \exp(i4\pi w_a
\pi T \tau)$, would occur at $T=T_c$ in the homogeneous case. Since the typical
soliton-antisoliton separation $L_c$ is much smaller than $\xi(T)$, see discussion below
Eq.~(\ref{eq:solution_delta}), the non-stationary condensates do not appear.
\bibitem{snsjunction} In principle, the soliton-antisoliton pair may be viewed as an SNS junction, but
because its size, $L_c$ exceeds the coherence length in the interior, which is of the
order of $L_T$, the expected critical current is exponentially small.
\bibitem{prefactor} We neglect the magnetic field dependence of the
pre-exponential factor. We believe this is justified for $L_H \gg L_T,\xi(T) $.
\end{thebibliography}
\end{document}